\documentclass[prc, groupedaddress, twocolumn, nofootinbib]{revtex4-1}

\usepackage{graphicx}
\usepackage{amsmath}
\usepackage{multirow}
\usepackage{enumerate}
\renewcommand\Im{\operatorname{Im}}
\usepackage{srcltx}
\usepackage{ulem}


\begin{document}

\title{Final State Interactions Effects in Neutrino-Nucleus Interactions}
\author{Tomasz Golan}
\email{tgolan@ift.uni.wroc.pl}
\author{Cezary Juszczak}
\affiliation{{ }\\
Institute for Theoretical Physics, Wroc\l aw University\\
Plac Maxa Borna 9, 50-204 Wroc\l aw, Poland}
\author{Jan T. Sobczyk}
\affiliation{{ }\\
Fermi National Accelerator Laboratory, Batavia, Illinois 60510, USA}
 \altaffiliation[On leave from the ]{Institute for Theoretical Physics, Wroc\l aw University}%Li

\begin{abstract}
Final State Interactions effects are discussed in the context of 
Monte Carlo simulations of neutrino-nucleus interactions.  
A role of Formation Time is explained and several models describing 
this effect are compared. Various observables which are sensitive 
to FSI effects are reviewed including pion-nucleus interaction and hadron yields in backward hemisphere. 
NuWro Monte Carlo 
neutrino event generator is described and its ability to understand 
neutral current $\pi^0$ production data in $\sim 1$~GeV neutrino flux experiments is demonstrated.
\end{abstract}


\begin{large}\end{large}

\maketitle

%%%%%%%%%%%%%%%%%%%%%%%%%%%%%%%%%%
%%%%%%%%%%%%%%%%%%%%%%%%
\section{Introduction}
%%%%%%%%%%%%%%%%%%%%%%%
%%%%%%%%%%%%%%%%%%%%%%%%%%%%%%%%%

New generation of neutrino oscillation parameters measurements require a good 
knowledge of neutrino-nucleus cross sections. Experimental data analysis is always
based on predictions from Monte Carlo (MC) event generators \cite{gallagher}. 
%It is important that the theoretical models describing neutrino interactions allow for a MC implementation: 
%all the particles in the final state must have well determined
%momenta and the code must be efficient enough to produce large numbers of events in a reasonable time. 
%Both conditions are nontrivial to satisfy.
In the $1$~GeV energy region, characteristic for several oscillation 
experiments (MINOS, T2K, MiniBooNE, NOvA) the use of the
Impulse Approximation (IA) picture \cite{artur_ia} in which neutrinos scatter on 
individual quasi-free nucleons is well justified. 
In this picture any neutrino-nucleus interaction becomes a
two-step process: (i) the  primary scattering on a bound nucleon, and 
(ii) Final State Interactions (FSI) affecting the hadrons produced at the step (i). 
The FSI contribute significantly to the systematic 
errors in  neutrino oscillation measurements so it is 
important to develop models to describe them better 
and also to understand the models'
limitations \cite{patrick}.

MC codes used in major neutrino oscillation experiments (FLUKA 
\cite{fluka}, NUANCE \cite{nuance}, NEUT \cite{neut}, GENIE \cite{genie}) in their description of 
FSI effects rely on the model of intra-nuclear cascade (INC) 
\cite{metropolis}. It is 
a semi-classical approach in which some quantum effects can also be 
incorporated (Pauli blocking, formation time (FT), nucleon correlations). 
Theoretical arguments for the applicability of the cascade model go back to 
the works of Glauber \cite{glauber}. More recently the investigation of the 
cascade model in the $\Delta$ resonance region was done 
in \cite{oset}. The model predictions agree with the experimental data for 
the pion-nucleus reaction cross sections, including the pion absorption. 

While the basic idea behind the models of FSI in the MC codes is always 
the same, numerical implementations are quite different reflecting 
priorities of particular neutrino experiments (target, detection technique etc). 

An important and not sufficiently understood ingredient in the INC models are Formation Time (FT) effects. 
On the most fundamental level
the FT is related to the Quantum Chromodynamics phenomenon called the color transparency 
(CT), proposed by Brodsky and Mueller \cite{CT}. For high enough 
four-momentum transfers a quark system is created with a small transverse size 
(point-like configuration - PLC) which is supposed to suppress hadrons re-interactions. 
As the typical size of the PLC is of the order of 
$1/|Q|$ \cite{pionT}, the CT effects are expected to be seen mostly at higher energies. 
Moreover, two-quark systems are more likely to create PLC than 
three-quark ones so the effect is expected to be larger for pions, than for 
nucleons. 

Independent phenomenological considerations \cite{stodolsky}, \cite{rantf} led to the
construction of approximate models of FT. As will be shown in Sect \ref{sec: ft} many evaluations 
of basic parameters which determine the size of FT effects have been proposed. It seems
important to study them explicitly in the context of neutrino measurements. For example, the FT effects are in the obvious
interplay with the pion absorption, reducing its probability in a non trivial momentum dependent way.

A validation of FSI models can be based  on any hadronic 
observables as all of them are FSI sensitive. Such observables include: distributions of numbers of reconstructed hadron tracks, 
spectra of hadrons in the final state, their
angular distributions etc. FSI models used in neutrino MC simulations can also be validated
on electro- and photo- nucleus observables. In the analysis of the NOMAD high energy neutrino scattering data 
\cite{nomad_ccqe} the introduction of FT was necessary 
to get the agreement with the measurements of backward moving protons and pions. 
It is interesting that the analysis 
of hadron-nucleus scattering data within INC models indicate that also at lower energies the 
FT effects can be quite important \cite{pion_INC}. 

In this paper FSI effects are modeled within 
the NuWro MC event 
generator \cite{nuwro}. NuWro covers neutrino energy range from a few hundreds MeV (the limit of 
applicability of the Impulse Approximation - IA) to several TeV. The code has flexibility to include Spectral Function \cite{spectral} 
formalism with sophisticated nuclear effects, as an 
alternative to the Fermi Gas model or an momentum dependent effective potential 
\cite{jns}. NuWro 
allows for comparisons to the data reported by experimental groups in the FSI 
effects included format.% \cite{morgan_dehradun}. 

\begin{figure}[tb]
 \includegraphics[width = \columnwidth]{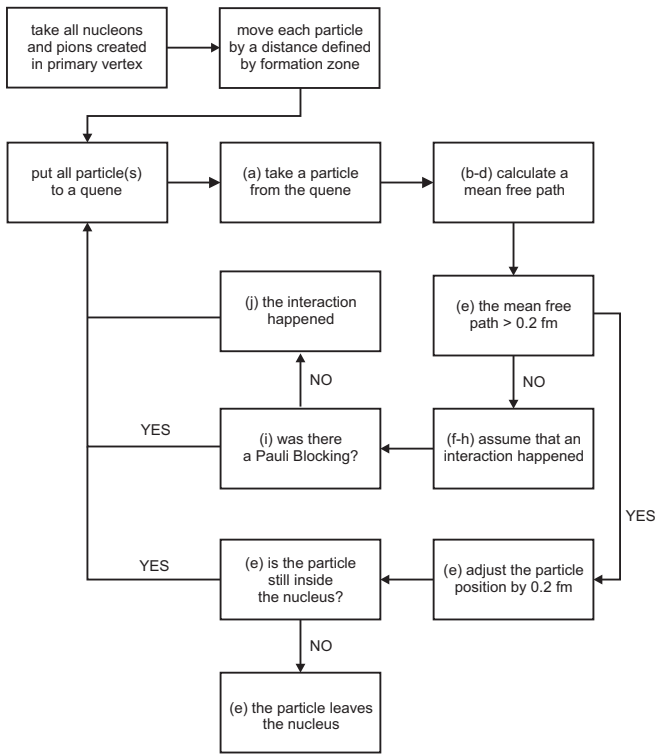}
\label{fig: cascade}
\caption{A block diagram of  NuWro INC algorithm.}
\end{figure}

We will consider a model of FT which is validated on the NOMAD backward moving pions data.
We will then discuss the NC $\pi^0$ 
production data and see how important the FT effects are for 
the understanding of the experimental data. NC 1$\pi^0$ is a very important process because it is a background 
to $\nu_\mu \rightarrow \nu_e$ oscillation search in water Cherenkov 
detectors: it can happen that one of two photons from $\pi^0$ decay remains 
undetected and the other is reconstructed as an electron.
NC 1$\pi^0$ production is also a very useful reaction to validate FSI models in the 1 GeV energy region. 
It is very sensitive to pion absorption and it is important to investigate how relevant are 
FT effects making the nuclear environment more transparent for pions produced inside nucleus. 

The paper is organized in the following way. In Sect.\ 2 a general 
description of the NuWro MC model is given. In Sect.\ 3 the NuWro FSI 
model based on the theoretical approach of Oset \cite{oset} is described. Several tests 
are reported showing a good agreement with the original
numerical implementation. Sect.\ 4 contains a sumary of various ways to 
model the FT. Various
approaches considered in the context of neutrino interactions 
and parameters used in theoretical computations and in MC codes are discussed. 
In Sect 5 the NuWro predictions are compared with the
NC 1$\pi^0$ production data and the significance of the FT
effects is discussed. Our conclusions are contained in Sect 6. 
%%%%%%%%%%%%%%%%%%%%%%%%%%%%%%%%%
%%%%%%%%%%%%%%%%%%%%%%%%%%%%%%%%
\section{NuWro}
%%%%%%%%%%%%%%%%%%%%%%%%%%%%%%%%%
%%%%%%%%%%%%%%%%%%%%%%%%%%%%%%%%%

NuWro is a neutrino event generation software developed at the 
Wroc\l aw University \cite{nuwro}. 
The main motivation for the NuWro authors was to have a tool to investigate the impact of 
nuclear effects on directly observable quantities with all the FSI effects included. 
Since 2005 it evolved
into a fairly complete neutrino interactions modeling tool. Its basic architecture is similar to 
better known MCs like NEUT or GENIE. 
All  major 
neutrino-nucleus interaction channels 
are implemented and the commonly used relativistic Fermi Gas (FG) model 
is for certain nuclei replaced with the more realistic Spectral 
Function model \cite{spectral}.
The NuWro FSI code has recently been updated by implementing the Oset model 
\cite{oset} of effective pion-nucleon cross sections and several options for the FT. 
Other upgrades include: parameterization 
of the multipion production cross section in pion-nucleon collisions based on the 
available data and the implementation of angular distributions in elastic and charge exchange 
pion-nucleon scattering based on the SAID model \cite{said}. 

With the inclusion of realistic beam models and a 
detector geometry module NuWro is becoming a fully-fledged MC event generator ready for use in neutrino experiments.

%%%%%%%%%%%%%%%%%%%%%%%%%%%
\subsection{Interactions} 
%%%%%%%%%%%%%%%%%%%%%%%%%%%

In NuWro there are four basic dynamic channels: quasi-elastic (QEL), resonance (RES), more inelastic (DIS) and 
coherent pion production (COH), each can be either in the charged current (CC) or in the neutral current (NC) mode.
The eight channel/mode combinations can be individually enabled or disabled.

For each (but the coherent) channel a particular nucleon which will take
part in the interactions is picked up
with nuclear matter density used as the probability density. Its momentum 
is chosen
from a ball with the radius set to the Fermi momentum (or the
local Fermi momentum calculated for that density in the case of LDA)
or obtained as a draw from the Spectral Function.

%%%%%%%%%%%%%%%%%%%%%%%
\subsubsection{QEL}
%%%%%%%%%%%%%%%%%%%%%%%

The CC quasi-elastic and NC elastic reactions are handled by the QEL channel.
It uses the standard Llewellyn Smith formulae \cite{llewelyn} with several options for the vector form factors (dipole, BBA03 \cite{BBA03}, BBBA05 \cite{BBBA05}, Alberico et al \cite{alberico_graczyk}).

The global and local relativistic FG models or SF approach are typically used but 
the kinematics based on the momentum dependent nuclear potential \cite{jns} is also available.
Currenly,  the Spectral Functions  for 
carbon, oxygen, argon, calcium and iron are implemented in NuWro according 
to the tables obtained from Omar Benhar or calculated in \cite{ankowski_sobczyk}. 
In the SF mode the de Forest kinematical prescription \cite{deforest} is used.

\begin{figure}
\begin{center}
 \includegraphics[width=\columnwidth]{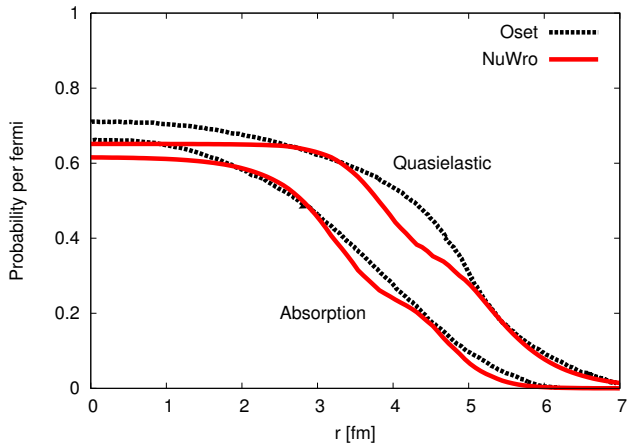}
\caption{Probability (per fermi) of microscopic pion-nucleon interactions as a function of a distance from the centre of an iron nucleus. 
Pion kinetic energy $T_k = 165$~MeV. The Oset model results are taken from \cite{oset}.  }
\label{fig: probasr}
\end{center}
\end{figure}

\begin{figure}
\begin{center}
 \includegraphics[width=\columnwidth]{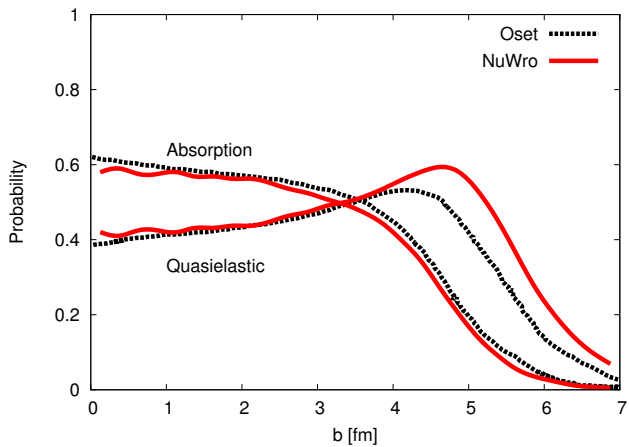}
\caption{Probability of macroscopic quasi-elastic or absorption interactions as a function of an impact parameter $b$ for $\pi^+ {}^{40}$~Ca 
scattering with pion kinetic energy $T_k = 180$~MeV. The Oset model results are taken from \cite{oset}. }
\label{fig: probasb}
\end{center}
\end{figure}

%%%%%%%%%%%%%%%%%%%%%%
\subsubsection{RES}
%%%%%%%%%%%%%%%%%%%%%

The RES channel is defined as $W<1.6$~GeV, where $W$ is the invariant hadronic mass. The dominant 
contribution comes from the single  pion production mediated by the $\Delta (1232)$ resonance according 
to the model \cite{delta}. Axial form-factors are taken from the reanalysis of the ANL and BNL bubble chambers data \cite{warszawa_wroclaw_ff}. Not using the standard resonance production Rein-Sehgal model \cite{resrein} to describe a contribution from higher 
resonances is justified by the quark-hadron duality hypothesis \cite{GJSduality} and by the fact that higher resonances cannot be 
separated in lepton-nucleus scattering. The non-resonant background is modeled as a fraction of the DIS contribution for 
$W\in (1.3, 1.6)$~GeV scaled so as the passage to the pure DIS channel be smooth.

%%%%%%%%%%%%%%%%%%%%%%%
\subsubsection{DIS}
%%%%%%%%%%%%%%%%%%%%%%

The DIS channel is defined as $W>1.6$~GeV. The total cross sections are evaluated using the Bodek-Yang prescription \cite{bodek_yang}. 
The Pythia6 hadronization routine is called for specific quark configurations 
\cite{sartogo} to allow their meaningful use also in the small $W$ region down to $1.2$~GeV.

%%%%%%%%%%%%%%%%%%%%%%
\subsubsection{COH}
%%%%%%%%%%%%%%%%%%%%%%%

The coherent pion production is implemented using the Rein-Sehgal model \cite{cohrein} with lepton mass corrections.

%%%%%%%%%%%%%%%%%%%%%%%%%%%%%
%%%%%%%%%%%%%%%%%%%%%%%%%%%%%
\section{NuWro FSI model}
%%%%%%%%%%%%%%%%%%%%%%%%%%%%%%%%%%%%
%%%%%%%%%%%%%%%%%%%%%%%%%%%%%

The NuWro FSI effects are described in a framework of the INC model \cite{metropolis}. 
The neutrino interaction point 
is selected inside nucleus according to the nuclear matter density. 
All secondary hadrons propagate through nucleus and can interact with 
nucleons inside. In the code the $0.2$~fm step length is assumed. 
For smaller values of the step the results remain the same, 
only running time increases. Between the collisions hadrons are assumed to be 
on-shell and move in straight lines. At each point of their path it is
decided if there was an interaction or not. This is done based on an effective 
cross section model. The generic re-interaction algorithm is independent on 
the dynamics used, also different models of nuclear density can be used. 
The particular dynamics is taken from the Oset model \cite{oset}, as it has 
solid theoretical foundations. The model is supposed to work well in the 
most important $\Delta$ region for pion kinetic energies in the range 
$85-350$~MeV. Outside this region the cross sections are obtained from
parameterizations  of  the available pion-nucleon cross section data. 

\begin{figure}
\begin{center}
\includegraphics[width = \columnwidth]{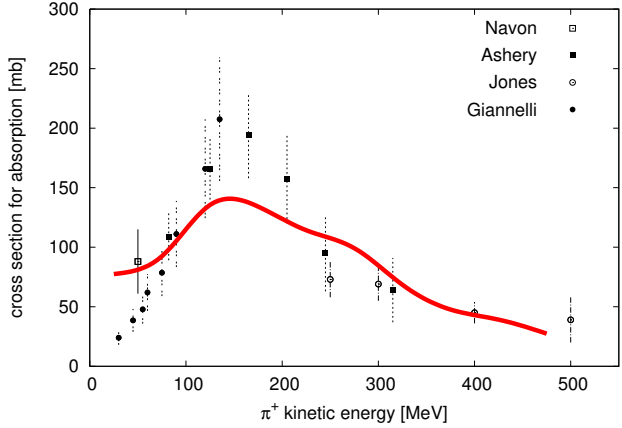}
\caption{$\pi^+\ ^{12}$C absorption cross section. The data points are taken from: 
Ashery \cite{ashery}, Navon \cite{navon}, Jones  \cite{jones} and Giannelli \cite{gianelli}. 
The solid line shows NuWro predictions. }
\label{fig: absorption}
\end{center}
\end{figure}

\begin{figure}
\begin{center}
\includegraphics[width = \columnwidth]{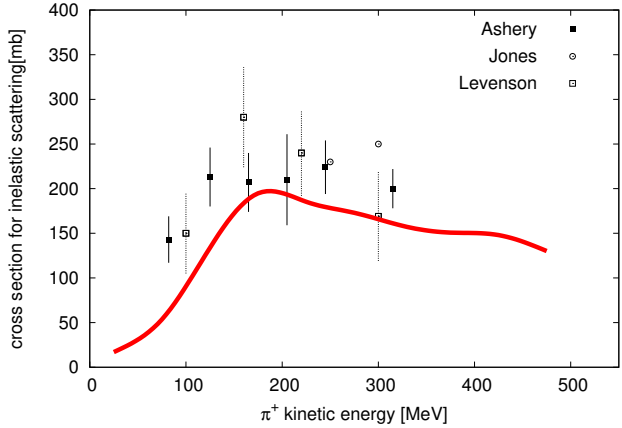}
\caption{$\pi^+\ ^{12}$C inelastic cross section. The data points are taken from: 
Ashery \cite{ashery}, Jones \cite{jones}, and Levenson \cite{levenson}. The solid line shows NuWro predictions.}
\label{fig: inelastic}
\end{center}
\end{figure}

\begin{figure}
\begin{center}
\includegraphics[width = \columnwidth]{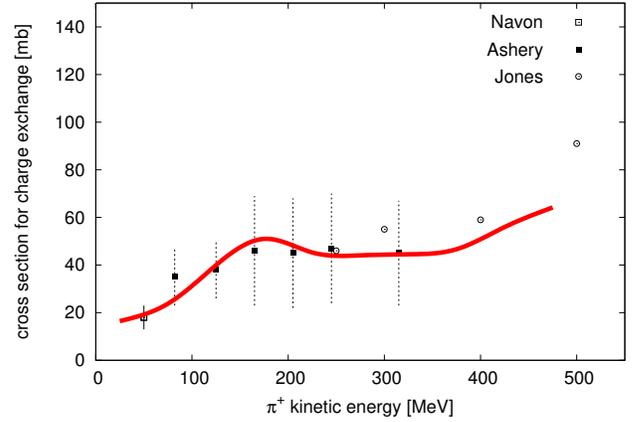}
\caption{$\pi^+\ ^{12}$C charge exchange cross section. The data points are taken from: Ashery \cite{ashery}, Navon \cite{navon}, 
and Jones \cite{jones}. The solid line shows NuWro predictions.}
\label{fig: scx}
\end{center}
\end{figure}

The basic FSI scheme consists in putting nucleons and pions produced in the
primary and also in secondary interactions to a 
queue and repeating the following until the queue gets empty:

\begin{enumerate}[(a)]

\item take a particle from the queue,

\item examine the nucleus density at its position,

\item calculate the mean free path,

\item probe the exponential distribution for the particle paths,

\item if the selected path is bigger then $0.2$~fm 
adjust the particle position by $0.2$~fm,

	\begin{itemize}

	\item if the particle is still in the 
         nucleus put it at the end of the queue,

         \item if the particle is outside the nucleus 
         put it to the list of outgoing particles,
	\begin{itemize}
\item[$1^o$] nucleons kinetic energy is diminished by the value of the potential: 

$$V = \sqrt{M^2 + k_f^2} - M + 8\ MeV$$
where $k_f$ is the Fermi momentum and its momentum adjusted so that it remains on the mass shell.

\item[$2^o$] if nucleons kinetic energy is smaller then 
$V$ the step $1^o$ cannot be completed. The nucleon is assumed to be 
unable to leave nucleus. It is reinserted to the nuclear matter and its kinetic energy contributes to the nucleus excitation energy,

         \end{itemize}
         \end{itemize}

\item if the selected path is smaller that $0.2$~fm 
assume that interaction with nuclear matter happened at this very 
place

\item probe the target nucleon momentum from the Fermi ball with the
local Fermi momentum calculated from the density at that point
\item select the type of interaction and generate the kinematics %an interaction 
\item check if none of the resulting nucleons 
is Pauli blocked; in the case of Pauli blocking forget 
the interaction, reinsert particle to the queue at the failed 
interaction point (Pauli blocking effects can be also included by means of
increased values of the 
mean free paths and then this step of the algorithm must be skipped)

\item  if the interaction was not Pauli blocked, all the particles in the final state are put to the end of the queue. 

\item if FT/FZ effects apply also to secondary interactions (this is model dependent) the particles positions are accordingly adjusted.

\end{enumerate}

The nucleus radius is defined as a distance from the center,
where the density is smaller by a factor of $10^4$ than the maximal one.

As a result of some nucleons joining the INC, 
the nuclear matter density is reduced but the shape of the density profile is assumed to be unchanged. 

%%%%%%%%%%%%%%%%%%%%%%%%%%%%%%
\subsection{The Oset model}
%%%%%%%%%%%%%%%%%%%%%%%%%%%%%%%

On the microscopic level the Oset model includes the quasi-elastic pion-nucleon reaction 
(including the charge exchange channel) and the pion absorption with two- and three-body absorption mechanisms. 
The interaction probability per time unit is:

\begin{equation}
 Pdt = - \frac{1}{\omega}\Im (\Pi )dt = -2 \Im (V_{opt})dt
\end{equation}

where $\omega$ is the pion energy, $\Pi$ is the pion self-energy, 
and $V_{opt}$ is the optical potential.

In the simplest case of $\pi^+ p \rightarrow \pi^+ p$ p-wave scattering calculations lead to the result:

\begin{equation}
 P = \frac{1}{\omega} \frac{2}{3} \left(\frac{f^*}{m_\pi}\right)^2 q_{c.m.}^2 |G_\Delta (q)|^2 \frac{1}{2}\Gamma \rho_p
\label{piplusp}
\end{equation}

where $f^*$ is $\pi N \Delta$ coupling constant $\left(f^{*2}/4\pi = 0.36\right)$, $m_\pi$ is the pion mass, 
$q_{c.m.}$ is the pion momentum in the centre of mass system, $G_\Delta$ is $\Delta$ propagator, $\Gamma$ its width and $\rho_p$ is
proton density.

An important in-medium effect is the $\Delta$ self-energy. Its imaginary part 
can be parameterized as \cite{oset_salcedo}:

\begin{equation}
 \Im\Sigma_\Delta (\omega) = -\left[C_Q(\rho/\rho_0)^\alpha + C_{A2}(\rho/\rho_0)^\beta + C_{A3}(\rho/\rho_0)^\gamma\right]
\label{selfenergy}
\end{equation}

The $\Delta$ width is modified $\frac{1}{2}\tilde\Gamma \rightarrow 
\frac{1}{2}\tilde\Gamma - \Im\Sigma_\Delta$, changing the $\Delta$ 
propagator and producing extra terms in Eq. (\ref{piplusp}), 
proportional to functions $C$'s present in Eq.(\ref{selfenergy}). 
The term proportional to $C_Q$ corresponds to higher order 
quasi-elastic scattering and the terms with $C_{A2}$ and 
$C_{A3}$ correspond to two- and three-body absorption. $\rho$ is the 
nuclear matter density and $\rho_0=0.17fm^{-3}$ is the normal density.

The final expression for the interaction probability in the nuclear matter is:

\begin{equation}
 P = \frac{1}{\omega}  \int \frac{d^3k}{(2\pi)^3} n(\vec k) \frac{2}{3} \left(\frac{f^*}{m_\pi}\right)^2 q_{c.m.}^2 |G_\Delta (q+k)|^2 \frac{1}{2}\tilde{\Gamma}(q+k)
\label{eq: prob}
\end{equation}
where $n(\vec k)$ is the occupation number for protons/neutrons.

The $\Delta$ self-energy depends strongly on the nuclear density and
the pion absorption is more likely to occur in 
the central part of the nucleus. 

Finally, one introduces improvements to the model 
coming from: the $\pi N$ interaction s-wave contribution, the real part of the optical 
potential and finite size effects.

%%%%%%%%%%%%%%%%%%%%%%%%%%%%%%%%%%%%%%%%%%%%%%%%%%%%%%%
\subsection{NuWro implementation of the Oset model}
%%%%%%%%%%%%%%%%%%%%%%%%%%%%%%%%%%%%%%%%%%%%%%%%%%%%%%%

The Oset model is implemented in NuWro by means of 
tables containing the cross-sections as the functions of pion kinetic energy at various nuclear matter densities. 
Because finite size effects are not universal it was necessary to prepare tables for each isotope separately.

In the analysis of the performance of the cascade model one should distinguish microscopic (pion-nucleon) and macroscopic (pion-nucleus) reactions. 
Figs \ref{fig: probasr}-\ref{fig: probasb} 
show a comparison between our implementation and the original Oset 
model. Fig. \ref{fig: probasr} shows the inverse of 
mean free paths (or equivalently: interaction probabilities per fermi)
for two microscopic interactions as functions
of the distance from a nucleus center. In both cases contributions from pion-proton and pion-neutron reactions are added. There is a significant
dependence on the nuclear density: in the nucleus central region the absorption
probability is large. In the peripheral region the 
quasi-elastic scattering dominates overwhelmingly. Fig. \ref{fig: probasb} shows probabilities of 
macroscopic processes as functions of the impact parameter. 
One can see that for small values of the 
impact parameter 
the absorption is more likely than the quasi-elastic scattering.
One should remember that an incident pion can be absorbed also after one or 
more microscopic quasi-elastic scatterings.

\begin{table}
\begin{center}
\begin{tabular}{c||ccc||cccc}
  & \multicolumn{3}{c||}{$T_{\pi} = 85$~MeV} & \multicolumn{4}{c}{$T_{\pi} = 245$~MeV} \\ \hline
  &  n=1 & n=2 & n=3 & n=1 & n=2 & n=3 & n=4 \\ \hline
 Oset & 0.90 & 0.09 & 0.01 & 0.69 & 0.25 & 0.05 & 0.01 \\ 
 NuWro & 0.89 & 0.10 & 0.01 & 0.67 & 0.24 & 0.07 & 0.02 \\ \hline
\end{tabular}
\caption{Probabilities that macroscopic quasi-elastic process proceeds through n microscopic collisions. 
Oset model results are taken from \cite{oset}.}
\label{tab: pqe}
\end{center}
\end{table}

\begin{table}
\begin{center}
\begin{tabular}{c||ccc||cccc}
 & \multicolumn{3}{c||}{$T_{\pi} = 85$~MeV}& \multicolumn{4}{c}{$T_{\pi} = 245$~MeV} \\ \hline
 &  n=0 & n=1 & n=2 & n=0 & n=1 & n=2 & n=3 \\ \hline
 Oset & 0.81 & 0.17 & 0.02 & 0.37 & 0.41 & 0.17 & 0.04 \\
 NuWro & 0.87 & 0.12 & 0.01 & 0.41 & 0.37 & 0.16 & 0.05
\end{tabular}
\caption{Probabilities that pion absorption occurs after n$th$ quasi-elastic microscopic scatterings. 
Oset model results are taken from \cite{oset}.}
\label{tab: abs}
\end{center}
\end{table}

Tables \ref{tab: pqe}-\ref{tab: abs}
help to understand other aspects of $\pi^+ {\;}^{12}$C scattering. 
Again, we compare results from the original Oset paper with the
NuWro implementation. 
Table  \ref{tab: pqe} presents the probabilities that a macroscopic 
quasi-elastic event proceeds through $n$ collisions.  
Table \ref{tab: abs} contains the probabilities that absorption occurs exactly after $n$ microscopic quasi-elastic pion scatterings.
The results are shown for two values of incident pion kinetic energy:
$85$ and $245$~MeV. More energetic pions are likely to 
undergo several 
scatterings in which they loose a fraction of their energy until they fall into the absorption peak in the $\Delta$ region.

With a satisfactory agreement for microscopic ingredients of the 
Oset model, we present a comparison of the NuWro cascade model predictions
with experimental data for the $\pi^+\ ^{12}$C scattering (Figs \ref{fig: absorption} \textendash\ \ref{fig: scx}). 

One can distinguish the following macroscopic pion-nucleus reactions: elastic, charge exchange, absorption, 
and inelastic scattering.  
We do not show the results for the double charge exchange reaction 
because the cross section is very small. For larger energies the inelastic cross section
contains a pion production component.

The cross section measurement of the charge exchange and absorption processes is straightforward. 
The inelastic cross section is obtained in the indirect way as:

\begin{equation}
 \sigma_{inel} = \sigma_{total} - \sigma_{elastic} - (\sigma_{absorption} + \sigma_{CEX}),
\end{equation}

where the elastic pion-nucleus cross section contribution is 
evaluated based on theoretical and experimental arguments (for the details see \cite{ashery}).

The NuWro predictions are obtained in the standard way by arranging  
a homogeneous flux of pions and counting the particles 
in the final
state assuming that at least one microscopic interaction took
place. In the simulations on carbon the impact parameter is limited to $b<6.5$~fm. We checked that with larger $b$ 
values the evaluated cross sections do not change. In the MC simulations one cannot model elastic pion-nucleus reaction. The sum over all possible interaction channels gives the pion-nucleus reaction cross section.

Figs \ref{fig: absorption}-\ref{fig: scx} show  that the NuWro predictions are in a good 
agreement with the data.

FT effects can be also used in the secondary interaction but their 
impact on the final results is very small. 

%%%%%%%%%%%%%%%%%%%%%%%%%%%%%%%%%%%%%%
\section{Formation Time/Zone}
\label{sec: ft}
%%%%%%%%%%%%%%%%%%%%%%%%%%%%%%%%%%%%%%

%%%%%%%%%%%%%%%%%%%%%%%%%%%%
\subsection{Generalities}
%%%%%%%%%%%%%%%%%%%%%%%%%%%%

\begin{figure}[tbh]
\begin{center}
 \includegraphics[width = \columnwidth]{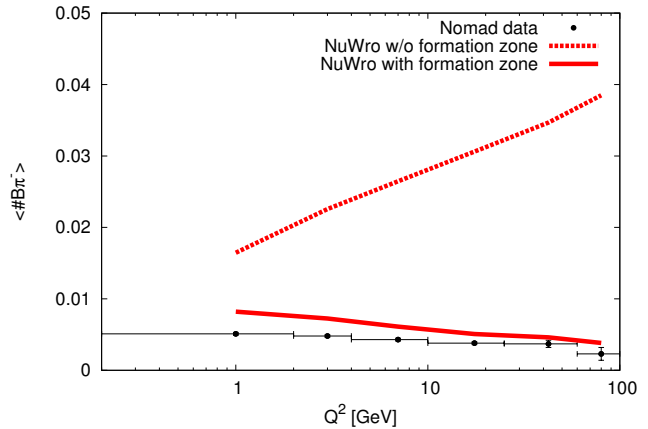}
 \caption{Average number of backwards going pions as a function of 
$Q^2$ in the NOMAD experiment.}
 \label{fig: nomad}
\end{center}
\end{figure}

The concept of the Formation Time/Formation Zone (FT/FZ) was introduced by Landau and Pomeranchuk \cite{landau_pomeranchuk} 
in the context of multiple scattering of electrons passing through a layer of material.
In the LAB frame the FT is given as:

\begin{equation}
 t = \frac{E}{k\cdot p}
\label{eq: lp}
\end{equation}

where $p^\mu = (E, \vec p)$ and $k^\mu = (\omega, \vec k)$ 
are four-momenta of the electron and the emitted photon respectively. 
$t$ has the interpretation of the minimal time necessary for a photon to be created.

The idea of FT was applied to hadron production by Stodolsky  \cite{stodolsky}, who considered the multiproduction of 
mesons by protons passing through a nucleus. In the Eq.~\ref{eq: lp} he replaced the electron by a 
projectile hadron with a four-momentum $p_0^\mu = (E_0, 0, 0, 
\sqrt{E_0^2 - M_0^2})$ and the photon by a secondary hadron with 
a four-momentum $p^\mu = (E, \vec p_T, \sqrt{E^2 - p_T^2 - M^2})$ obtaining:

\begin{eqnarray}
t\rightarrow t_f & = & \frac{E_0}{EE_0 - \sqrt{E^2 - p_T^2 - M^2}\sqrt{E_0^2 - M_0^2}} \nonumber \\ & = & \frac{1}{E\left(1 - \sqrt{1 - \frac{\mu_T^2}{E^2}}\sqrt{1 - \frac{M_0^2}{E_0^2}}\right)}
\end{eqnarray}

where $\mu_T$ is the transverse mass defined as 
$\mu_T^2 \equiv M^2 + p_T^2$. 

For higher energies $E \gg \mu_T$, $E_0\gg M_0$ and

\begin{equation}
t_f \approx \frac{2E}{(M_0x)^2 + \mu_T^2}
\label{eq: stod} 
\end{equation}

where $x = \frac{E_0}{E}$. Rantf \cite{rantf} argued that a further simplification $x\approx 0$ is usually well 
justified and finally in the LAB frame:

\begin{equation}
t_f \approx \frac{2E}{M^2 + p_T^2}
\label{stod_lab} 
\end{equation}

and in the hadron's rest frame:

\begin{equation}
t_{f,rest} \approx \frac{2M}{M^2 + p_T^2}.
\label{stod_rest} 
\end{equation}

Inspired by this expression Rantf postulated another formula for the FT in the hadron rest frame. 
He kept the basic relativistic character of the FT but introduced an arbitrary parameter $\tau_0$ to control its size:

\begin{equation}
\tau_{rest}=\tau_0 \frac{M^2}{M^2 + p_T^2}.
\label{eq: ranft_final} 
\end{equation}

The FT defined in  Eq. \ref{eq: ranft_final} was implemented in the MC event generator 
DPMJET \cite{dpmjet} which later became a part of the FLUKA code and was used by the NOMAD collaboration \cite{nomad}. 
In the DPMJET cascade model the FT is applied to hadrons resulting from all the interaction modes: QEL, RES and DIS. 
Following the ideas of Bialas \cite{bialas} in DPMJET values of FT are sampled from the exponential distribution.

The FT played an important role in the NOMAD analysis of the CCQE events \cite{nomad_ccqe}. They populate 
mainly one- and two- tracks samples.  A change of $\tau_0$ modifies the MC predictions for the size of both samples: 
an increase of FT makes an impact of the FSI effects on the ejected protons smaller and they are more 
likely to have larger momentum with increased probability of being detected and populating the two-track sample of events. 
By adjusting the size of the formation time the values of $M_A$ calculated independently from either of 
the two samples of events became almost identical. 

In the above estimations of the FT effect several assumptions were made which are not necessarily valid at lower energies. 
This is taken into account in the more recent low energy FLUKA cascade model, called PEANUT \cite{peanut}. 
In the case of the QEL reaction the concept of coherence length (CL) was proposed to substitute the FT effect. 

Derivation of the CL is based on the uncertainty principle arguments: 
Let $p^\mu$ be the outgoing nucleon four-momentum
and $q^\mu = (\omega, \vec q)$ the four-momentum transfer, both in the LAB frame. 
Because $p\cdot q$ is a Lorentz scalar, one can calculate $\tilde\omega$ ($\tilde{}$\ denotes quantities calculated in the nucleon rest frame),
the energy transfer in the final nucleon rest frame:

\begin{equation}
 |p\cdot q| = |\tilde p \cdot \tilde q| = |\tilde\omega M| \Rightarrow |\tilde\omega| = \frac{|p\cdot q|}{M}
\end{equation}

From the uncertainty principle $\tilde\omega$ can be used to 
estimate the reaction time in the nucleon's rest 
frame and then in the LAB frame as well. Within that time the
nucleon is assumed to be unable to re-interact \cite{coh_length}:

\begin{equation}
 t_{CL, rest} = \frac{M}{|p\cdot q|} \qquad t_{CL} = \frac{E}{|p\cdot q|},
\label{eq: cohl}
\end{equation}

Surprisingly, the Landau-Pomeranchuk formula Eq. \ref{eq: lp} is reproduced. 

Among other approaches to give a quantitative evaluation of the FT effects one should mention
the SKAT parameterization of the LAB frame FZ \cite{baranov}:

\begin{equation}
l_{SKAT} = \frac{|\vec p|}{\mu^2}.
\label{eq: skat}
\end{equation}

The value of the free parameter was found to be 
$\mu^2 = 0.08 \pm 0.04$~GeV$^2$ based on the experimental data for the
multiplicity of low momentum ($300$~MeV/c $< p < 600$~MeV/c) protons. 
This value of $\mu^2$ agrees also with the 
analysis of the momentum distribution of negatively charged mesons in the region $p < 3$~GeV/c. 

The SKAT formula can be translated to the following value of FT:

\begin{equation}
t_{SKAT} = \frac{E}{|\vec p|}l_{SKAT} = \frac{E}{\mu^2}, \qquad
t_{SKAT, rest} = \frac{M}{\mu^2}.
\label{skat_FT}
\end{equation}

Compared to the Rantf formula (Eq. \ref{eq: ranft_final}) the SKAT parameterization  corresponds to $p_T=0$ but it also introduces a 
scale proportional to the hadron mass:
$ \tau_0 \leftrightarrow M/\mu^2$.  According to the SKAT parameterization FZ is identical for pions and nucleons with the same momentum. 
At $p\sim 1$~GeV/c FZ is expected to be $\sim 2.5$~fm, which is of the size of the carbon nucleus.  

In another approach to model FT effects, at a distance $z$ from the interaction point 
one postulates an effective (reduced) hadron-nucleon interaction cross section  \cite{eliseev}:

\begin{equation}
\sigma_{eff} (z) = \sigma_{free} \left(1-{\rm e}^{-zMm_0/|\vec p|}  \right)
\label{eliseev}
\end{equation}

with $m_0\approx 0.4$~GeV.

Similar description of the FT effects (reduction of the cross section) is 
used in the quantum diffusion model
\cite{farrar}:

$$\sigma^{eff}_{hN} (z) = \sigma^{}_{hN}$$
\begin{equation}
\times \left[ \left( \frac{z}{l_h} + \frac{<n^2k_t^2>}{Q^2}(1-\frac{z}{l_h})\right) \theta (l_h -z)
+ \theta(z-l_h)\right] ,
\label{diffusion}
\end{equation}

where $\sigma^{}_{hN}$ is the free hadron-nucleon cross section, $z$ is the distance from the interaction point, 
$k_t^2 \cong 0.35$~MeV/c is the average quark transverse momentum, $n=2, 3$ for pions and nucleons. The size of FT is determined by $l_h$ which can be evaluated to be

\begin{equation}
 l_h = 2p_h\left<\frac{1}{M_n^2 - M_h^2}\right>,
\end{equation}
where $p_h$ and $M_h$ denote hadron momentum and mass, and  $M_n$ is an intermediate state mass. The precise
value of $\Delta M^2 = M_n^2 - M_h^2$ is not known and is estimated to be between $0.25$ and $1.4$~GeV$^2$.
In \cite{cosyn} the values of $1$~GeV$^2$ and $0.7$~GeV$^2$ are used for protons and pions. 

In the parameterizations given in  Eqs (\ref{eliseev}) and (\ref{diffusion}) smaller effective 
cross section translates into a larger average distance to the first reinteraction point. 

In the case of pions produced via the $\Delta$ excitation and decay there is still another natural way to model the 
FT effect. In the INC picture we can treat
the $\Delta$ (like in  the GiBUU approach \cite{new_gibuu}) as a real particle propagating some distance 
before it decays. The $\Delta$ lifetime in its rest frame 
is equal $\frac{1}{\Gamma}$, with $\Gamma \approx 120$~MeV, so in the LAB frame one obtains:

\begin{equation}
 t_{\Delta } = \frac{E_\Delta}{M\Gamma}
\label{eq: delta}
\end{equation}

where $E_\Delta$ is the $\Delta$ energy in the LAB frame.

\begin{figure}
\begin{center}
\includegraphics[width = \columnwidth]{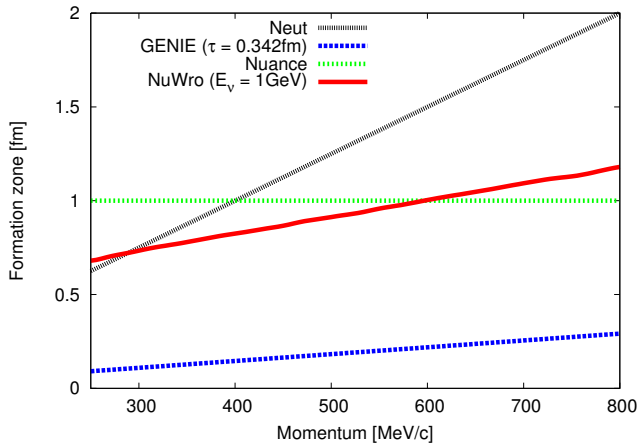}
\caption{Nucleon Formation Zone in the LAB frame as a function of its momentum.}
\label{fig: fzn}

\end{center}
\end{figure}

\begin{figure}
\begin{center}
\includegraphics[width = \columnwidth]{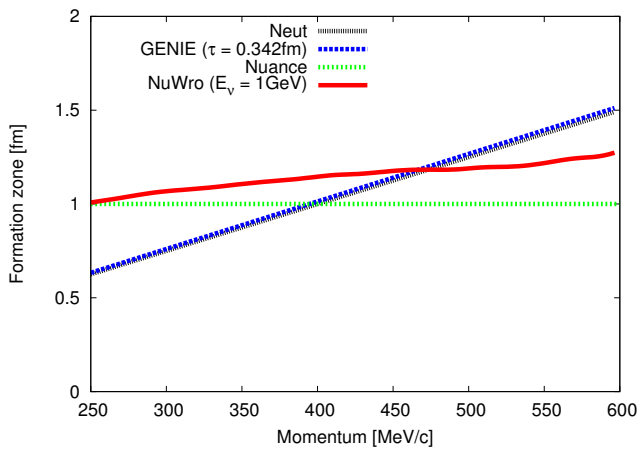}
\caption{Pion Formation Zone in the LAB frame as a function of its momentum }
\label{fig: fzp}
\end{center}
\end{figure}

We conclude that various approaches lead to similar expressions for the FT as far as the dependence on hadron momentum is 
concerned but numerical coefficient and the size of the effect can be quite different.

%%%%%%%%%%%%%%%%%%%%%%%%%%%%%%%%%%%%%%%%%%%%%%%%%%%%%%%
\subsection{FT models in MC event generators}
%%%%%%%%%%%%%%%%%%%%%%%%%%%%%%%%%%%%%%%%%%%%%%%%%%%%%%%

%In the  \ref{tab: fz} we collected available information about FT models in major neutrino MC event generators. 
Table \ref{tab: fz} summarizes available information about FT models in major neutrino MC event generators. 
\begin{table}[htb]
\begin{tabular}{l||c|c|c}
MC & QE & RES\footnote{Note that every MC has its own  slightly different definition of what does RES and DIS terms mean.}  & DIS \\ \hline
NEUT & -- & SKAT & SKAT \\
FLUKA & Coh length & Rantf & Rantf \\
GENIE & -- & -- & Rantf-like \\
NUANCE & 1 fm & 1 fm  & 1 fm
\end{tabular}
\caption{FT models in MC event generators}
\label{tab: fz}
\end{table}

NEUT uses the SKAT model both for RES and DIS \cite{hayato}.

FLUKA uses Eq. \ref{eq: cohl} for quasielastic scattering and Eq. \ref{eq: ranft_final} for other processes.

GENIE uses Eq. \ref{eq: ranft_final} but in a simplified form, neglecting $p_T$:

\begin{equation}
t_{\rm Genie} = \tau_0.
\end{equation}

GENIE assumes the value  $\tau_0=0.342 fm/c$ \cite{genie_official}. One can check that for the pions the SKAT formula is 
reproduced. GENIE applies FZ to DIS events and also to the non-resonant background events in the RES dynamics \cite{costas}.

NUANCE implemented an effective model in which the FZ is always equal to $1$~fm \cite{nuance_ft}.

%%%%%%%%%%%%%%%%%%%%%%%%%%%%%%%%%%%
\subsection{NuWro FT model}
%%%%%%%%%%%%%%%%%%%%%%%%%%%%%%%%%%%%

In NuWro the formation zone effects are implemented:

\begin{itemize}
 \item as coherence length (Eq. \ref{eq: cohl}) for quasielastic scatterings
 \item as $\Delta$ propagation (Eq. \ref{eq: delta}) for RES interactions
 \item using Ranft model (Eq. \ref{eq: ranft_final}) with some parameter $\tau$ for DIS 
\end{itemize}

There is a smooth transition between the last two models at $W \simeq  1.6$~GeV.

With the Ranft model (without the approximation $p_T = 0$) the average value of the 
FT depends both on neutrino energy and on hadron momentum. 
For a fixed value of the neutrino energy, lower hadron momenta typically correspond to larger values of the transverse momentum 
and smaller values of the FT. 

In order to fix the value of the parameter $\tau$ we analyze the NOMAD data for the backward moving pions.

%%%%%%%%%%%%%%%%%%%%%%%%%%%%%%%%%%%%%%%%%%%%%%%%%%%%%%%%%%%%%
\subsubsection{Comparison with NOMAD measurement of backward moving pions}
%%%%%%%%%%%%%%%%%%%%%%%%%%%%%%%%%%%%%%%%%%%%%%%%%%%%%%%%%%%%%

To fine tune our model of FZ we make use of the NOMAD experimental data \cite{nomad_multi}. 
The average neutrino energy in NOMAD is $\left<E_\nu\right> =24$ GeV and the target composition is dominated by 
carbon (64.30\%) and oxygen (22.13\%)  with small additions of other elements.

We focus on the pion data.
Our main observable is the average number of backward moving ($\cos\theta_{LAB}<0$) negative pions  B$\pi^-$ 
with the momentum $p_\pi$ between  $350$ and $800$ MeV/$c$, as a function of $Q^2$. This observable is
very sensitive to the FSI effects. Without FSI the number of backward moving pions would be very small because they appear mainly due
to nuclear reinteractions. Introduction of the FT makes the FSI effects smaller and reduces the number of B$\pi^-$. 

Simulations made for various  values of $\tau$ 
lead us to the conclusion that a good agreement with the data is obtained with 
$\tau\sim 8$~fm/c. 
Fig. \ref{fig: nomad} shows average numbers of backward moving $\pi^-$ 
reported by NOMAD, and predicted by NuWro with and without FZ, as a function of $Q^2$. 
In order to better understand the NuWro performance we analysed various ways in which B$\pi^-$ appear:

\begin{enumerate}
 \item pions are created in the primary vertex and undergo quasielastic scatterings during FSI
 \item pions are created during FSI in pion-nucleon interactions
\begin{enumerate}
 \item single pion production
 \item double pion production
 \item triple pion production
\end{enumerate}
 \item pions are created during FSI in nucleon-nucleon interactions
 \item there are more pion production processes during FSI.
\end{enumerate}

Contributions from the above scenarios to events with backward moving $\pi^-$ 
are listed in Table \ref{tab: nomadvivi}.

\begin{table}
\begin{tabular}{c|c|c}
Scenario & Without formation zone & With formation zone \\ \hline\hline
1 & 37.2\% & 83.7\% \\ \hline
2 & 43.3\% & 15.5\% \\ \hline
2a & 22.0\% & 8.1\% \\
2b & 15.6\% & 7.4\% \\
2c & 5.8\% & 0\% \\ \hline
3 & 2.7\% & 0.7\%\\ \hline
3a & 1.9\% & 0.6\% \\
3b & 0.8\% & 0.1\% \\ \hline
4 & 16.7\% & 0.1\%
\end{tabular}
\caption{Contribution to events with backward $\pi^-$ from different scenarios (description in text)}
\label{tab: nomadvivi}
\end{table}

Table \ref{tab: xbp} shows distributions of the number of $B\pi^-$ in a single event. 
The NuWro predictions (with and without the FZ) are compared with
the Nomad data \cite{nomad_multi}.

\begin{table}
\begin{tabular}{c|c|c|c}
 \multirow{2}{*}{$<\#B\pi>$} & \multirow{2}{*}{Data} & \multicolumn{2}{c}{NuWro} \\ \cline{3-4}
 & & Without FT & With FT \\ \hline
 0 & 939617 & 921048 & 937883  \\
 1 & 4238 & 22590 & 6126  \\
 2 & 164 & 375 & 8  
\end{tabular}
\caption{Contribution to B$\pi^-$ coming from events with 0, 1 or 2 pions}
\label{tab: xbp}
\end{table}

\begin{figure}[tb]
\begin{center}
\includegraphics[width = \columnwidth]{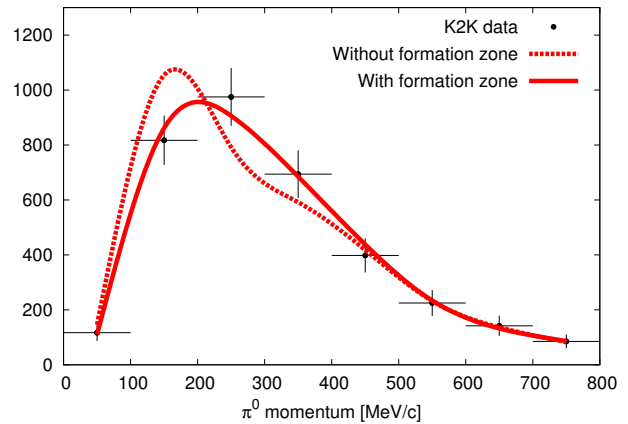}
\caption{K2K: NC $1\pi^0$ production as a function of $\pi^0$ momentum; the data and NuWro predictions are normalized to the same area.
}
\label{fig: k2k}
\end{center}
\end{figure}

\begin{figure}[tb]
\begin{center}
\includegraphics[width = \columnwidth]{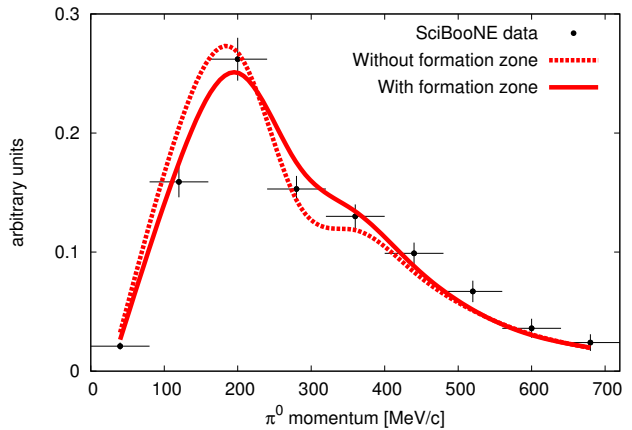}
\caption{SciBooNE: NC $\pi^0$ production as a function $\pi^0$ momentum; the data and NuWro predictions are normalized to the same area.}
\label{fig: sb}
\end{center}
\end{figure}

%%%%%%%%%%%%%%%%%%%%%%%%%%%%%%%%%%%%%%%%%%%%%%%%%%%%%%%%%%%%%
\subsubsection{Comparison with other MC event generators}
%%%%%%%%%%%%%%%%%%%%%%%%%%%%%%%%%%%%%%%%%%%%%%%%%%%%%%%%%%%%%

Figs \ref{fig: fzn} and \ref{fig: fzp} show the values of the FZ in NuWro  
compared to other 
MC neutrino event generators. It is interesting that in the case of pions various models of the FZ give very similar
results, while in the case of nucleons the differences are much larger.

The NuWro results are given for a specific neutrino energy ($E_\nu = 1GeV$). 
The FZ grows with $E_\nu$ due to the transverse momentum in the denominator, which goes to zero when hadron energy becomes higher. 

In NuWro the dependence of the FZ on the pion kinetic energy is very flat.

%%%%%%%%%%%%%%%%%%%%%%%%%%%%%%%%%%%
%%%%%%%%%%%%%%%%%%%%%%%%%%%%%%%%%%%
\section{Application: NC $1\pi^0$ production}
%%%%%%%%%%%%%%%%%%%%%%%%%%%%%%%%%%
%%%%%%%%%%%%%%%%%%%%%%%%%%%%%%%%%%

%%%%%%%%%%%%%%%%%%%%%%%%%%%%%%%%%%%%%%%%%%%%%%%%
\subsection{Free nucleon NC $\pi^0$ production}
%%%%%%%%%%%%%%%%%%%%%%%%%%%%%%%%%%%%%%%%%%%%%%%%%

The data for NC 1$\pi^0$ production cross section on a free nucleon target is 
very scarce.  The only such measurement so far was done in the Gargamelle bubble 
chamber. The target was in fact composed of C$_3$H$_8$ (90\%) and
CF$_3$Br but the FSI effects were subtracted according to the model in \cite{anp}. 

\begin{figure}[tb]
\begin{center}
\includegraphics[width = \columnwidth]{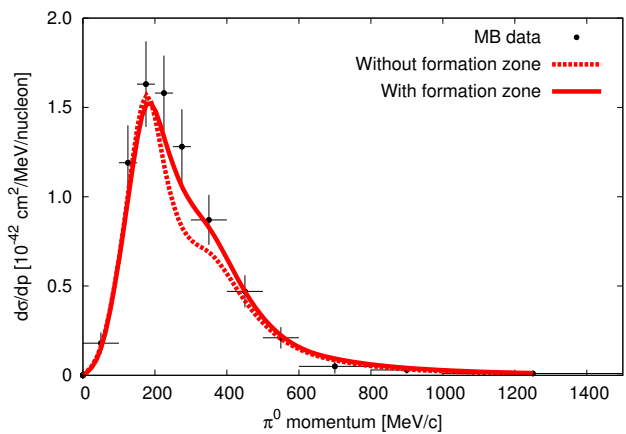}
\caption{MiniBooNE (neutrino mode): NC $1\pi^0$ production as a function $\pi^0$ momentum.}
\label{fig: mb}
\end{center}
\end{figure}

\begin{figure}[tb]
\begin{center}
\includegraphics[width = \columnwidth]{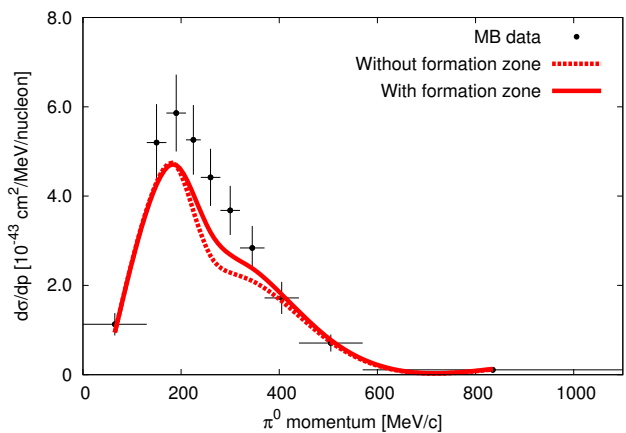}
\caption{MiniBooNE (anti-neutrino mode): NC $1\pi^0$ production as a function $\pi^0$ momentum.}
\label{fig: mba}
\end{center}
\end{figure}

In view of large uncertainties in the understanding of nuclear effects the results should be treated 
with some caution. We notice also that the data contain a contribution from the COH reaction. 
Originally, the results were presented as efficiency corrected relative production rates in several pion production 
channels \cite{krenz}. The data re-analysis was done in \cite{hawker}: information about neutrino flux
was taken into account and the cross sections estimations were done. 

Table \ref{tab: freepi} shows the experimental data from \cite{hawker} and NuWro predictions obtained for neutrinos of energy $2.2\,\,$GeV
on free nucleon target and also on nucleons bound in $^{12}C$, with Pauli blocking and 
Fermi motion effects taken into account but without FSI effects. We find the agreement to be 
satisfactory. It is also possible to compare the NuWro predictions and the data for 
relative contributions from RES and DIS reaction channels. In the case of $\nu p\rightarrow \nu p \pi^0$ reaction
the NuWro predictions for the RES:DIS ratio are: $78:22$ for free and $82:18$ for bound 
nucleons. The experimental data are $\sim 80:20$ (see Fig. 11 in \cite{krenz}).  

\begin{table}
\begin{tabular}{l|c|c|c}
  & \multicolumn{3}{c}{Cross section per nucleon ($\times 10^{-38} cm^2$)} \\ \hline
 \multirow{2}{*}{Channel} & \multirow{2}{*}{Data} & \multicolumn{2}{c}{NuWro} \\ \cline{3-4}
 & & free nucleon & bound nucleon \\ \hline
 $\nu_\mu p \rightarrow \nu_\mu p \pi^0$ & $0.13\pm 0.02$ & $0.15$ & $0.12$ \\
 $\nu_\mu n \rightarrow \nu_\mu n \pi^0$ & $0.08\pm 0.02$ & $0.17$ & $0.14$ \\
 $\nu_\mu p \rightarrow \nu_\mu n \pi^+$ & $0.08\pm 0.02$ & $0.13$ & $0.11$ \\
 $\nu_\mu n \rightarrow \nu_\mu p \pi^-$ & $0.11\pm 0.03$ & $0.14$ & $0.13$ \\
 $\nu_\mu n \rightarrow \mu p \pi^0$ & $0.24\pm 0.04$ & $0.38$ & $0.36$
\end{tabular}
\caption{Single pion production cross sections}

\label{tab: freepi}
\end{table}

%%%%%%%%%%%%%%%%%%%%%%%%%%%%%%%%%%%%%%%%%%%%%%%%%%%%%%%
\subsection{NC $1\pi^0$ production on a nucleus}
%%%%%%%%%%%%%%%%%%%%%%%%%%%%%%%%%%%%%%%%%%%%%%%%%%%%%%

\begin{figure}[tb]
\begin{center}
\includegraphics[width = \columnwidth]{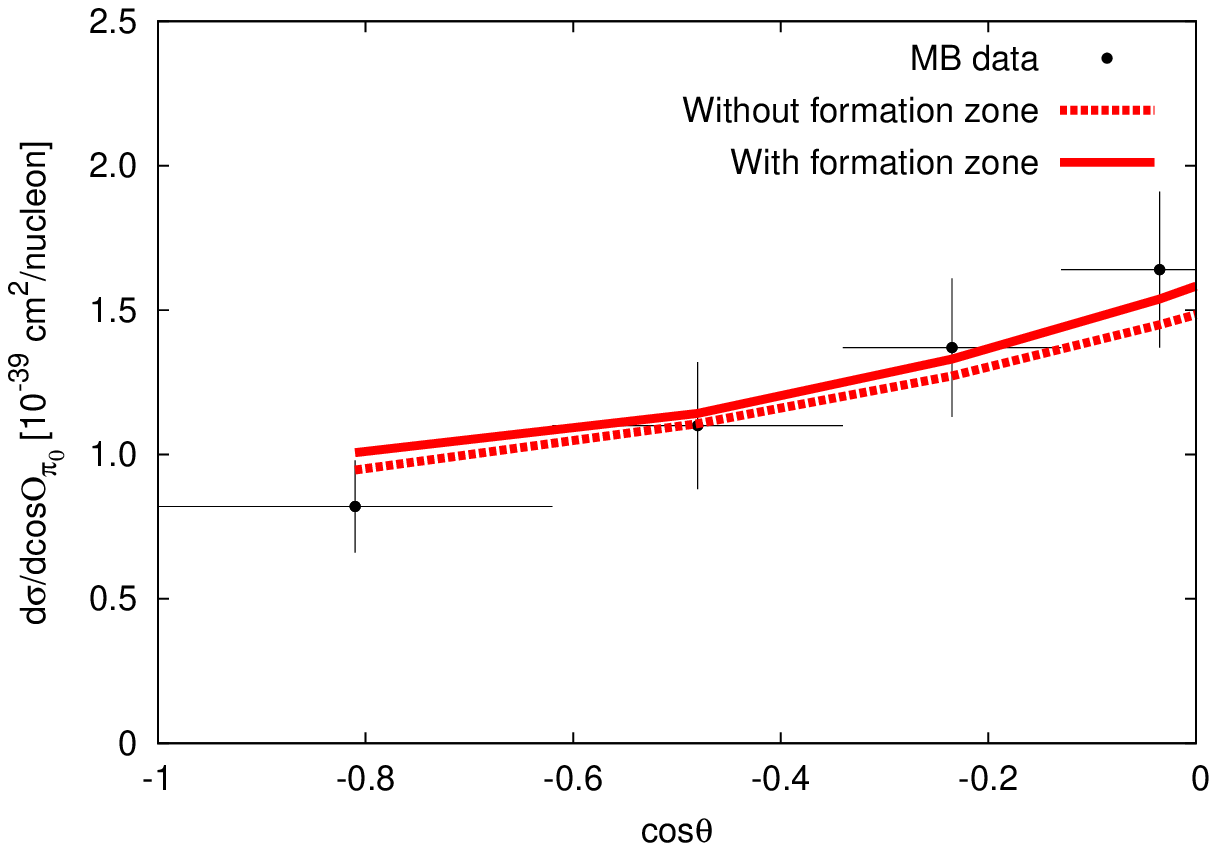}
\caption{MiniBooNE (neutrino mode): NC $1\pi^0$ production as a function $\cos\theta$.}
\label{fig: mb_angle}
\end{center}
\end{figure}

\begin{figure}[tb]
\begin{center}
\includegraphics[width = \columnwidth]{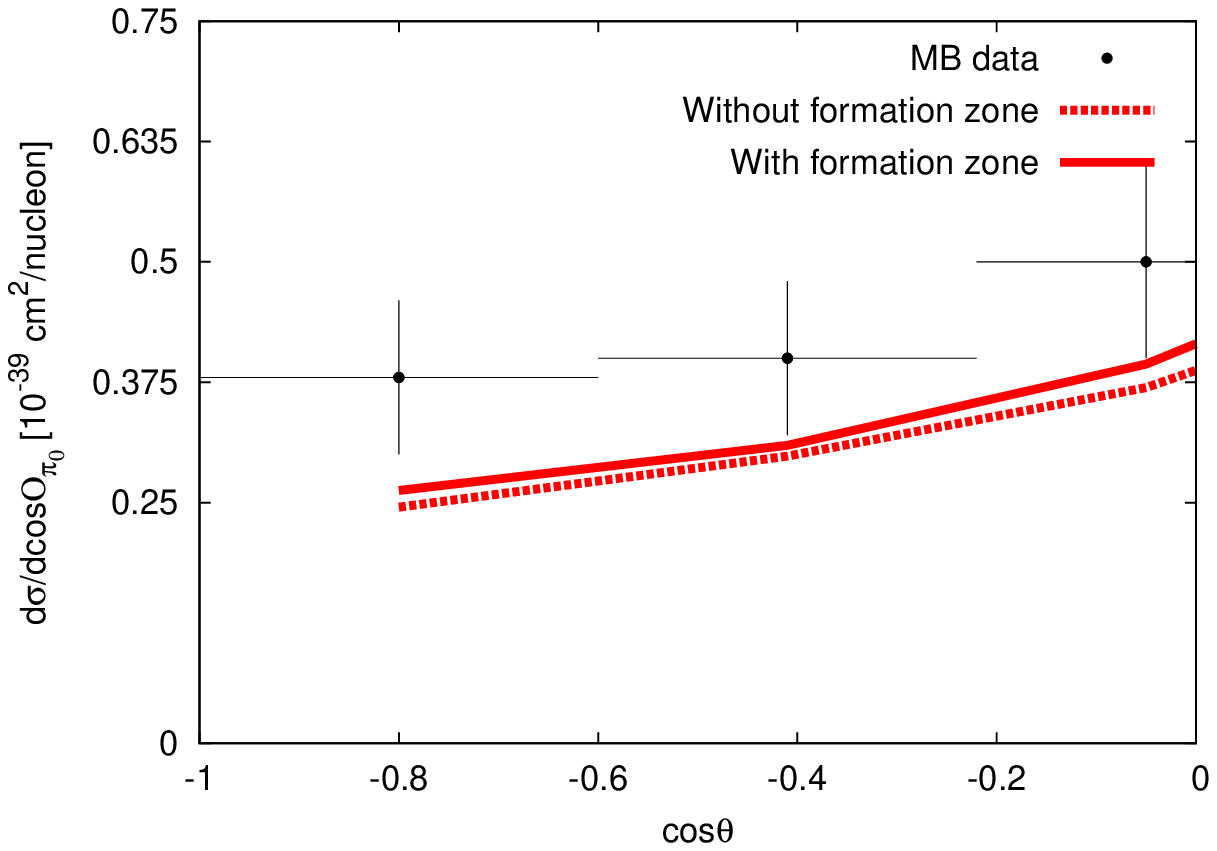}
\caption{MiniBooNE (anti-neutrino mode): NC $1\pi^0$ production as a function $\cos\theta$.}
\label{fig: mba_angle}
\end{center}
\end{figure}

\begin{figure}[tb]
\begin{center}
\includegraphics[width = \columnwidth]{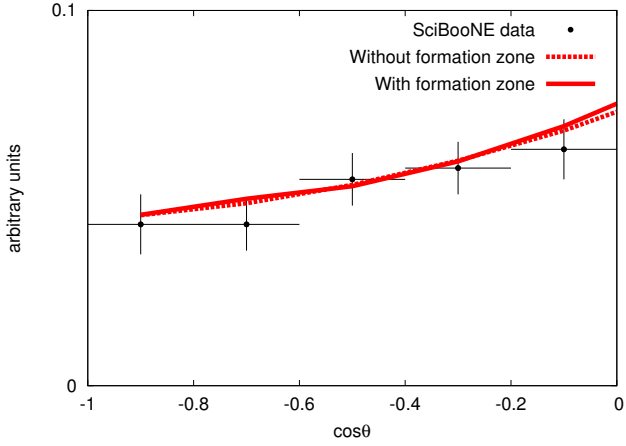}
\caption{SciBooNE: NC $\pi^0$ production as a function $\cos\theta$. }
\label{fig: sb_angle}
\end{center}
\end{figure}

The recent experimental data for NC $\pi^0$ production come from three experiments. 
Basic information about them is summarized in Table \ref{tab: ncpi0}.

\begin{table}[htb]
\begin{tabular}{c||c|c|c|c|c}
Experiment & Beam      & ${{\displaystyle \left<E_\nu\right>}\above0pt{\displaystyle \rm [GeV]}}$& Target &Normaliz.& Measurement  
\\ \hline
K2K \cite{k2k_ncpi0}       & $\nu_\mu$ & $1.30$& $H_2O$ &relative&${\displaystyle dN}/{\displaystyle dT_\pi}$ \\ \hline
\multirow{2}{*}{MB \cite{mb_ncpi0}}& \multirow{2}{*}{$\nu_\mu$} & \multirow{2}{*}{$0.81$} & \multirow{2}{*}{$CH_2$} 
& \multirow{2}{*}{absolute}& ${\displaystyle d\sigma}/{\displaystyle dT_\pi}$,  \\
& &  & & &  ${\displaystyle d\sigma}/{\displaystyle d\cos\theta_\pi}$ \\ \hline
\multirow{2}{*}{MB \cite{mb_ncpi0}}& \multirow{2}{*}{$\bar\nu_\mu$ }& \multirow{2}{*}{$0.66$ }& \multirow{2}{*}{$CH_2$} 
& \multirow{2}{*}{absolute}& ${\displaystyle d\sigma}/{\displaystyle dT_\pi}$\\
&  &  &  & &  ${\displaystyle d\sigma}/{\displaystyle d\cos\theta_\pi}$\\ \hline
\multirow{2}{*}{SciB \cite{sb_ncpi0}}& \multirow{2}{*}{$\nu_\mu$} & \multirow{2}{*}{$0.81$}& \multirow{2}{*}{$C_8H_8$ }& 
\multirow{2}{*}{relative}& ${dN}/{dT_\pi}$ \\
& & & & & ${\displaystyle dN}/{\displaystyle d\cos\theta_\pi}$ \\
\end{tabular}
\caption{Recent NC $\pi^0$ production measurements.}
\label{tab: ncpi0}
\end{table}

In the K2K and MiniBooNE (MB) experiments the signal was defined as  exactly one $\pi^0$ leaving the 
nucleus target and no other mesons in the final state. 
In the case of SciBooNE (SciB) the signal was defined as {\it at least} one $\pi^0$ in the final state, with possible other pions as well.

The experimental signal for 1$\pi^0$ production comes from: 
(i) single $\pi^0$ produced at the interaction point in the single pion production reaction; 
(ii) $\pi^0$  produced in double pion production reaction with other pion being absorbed; 
(iii) single $\pi^\pm$ production with charge exchange reaction $\pi^\pm\rightarrow \pi^0$ inside nucleus; 
(iv) primary quasi-elastic reaction with $\pi^0$ being produced due to nucleons re-interactions inside nucleus. 
For the 2$\pi^0$ production the number of possible scenarios is even bigger.

According to NuWro, most of the $1\pi^0$ signal events ($93-95\%$) come from the initial RES
single pion production reactions, see Table \ref{tab: ncpianal1}. In the case of the MB
antineutrino flux the contribution is the smallest, because the antineutrinos are on average less energetic. 
Also the impact of the FZ is in clear anti-correlation with the average flux energy.

Table \ref{tab: ncpianal2} enumerates what can happen to 
a $\pi^0$ produced in the primary vertex due to the FSI effects. 
One can see that pion absorption (the second row) reduces the number of NC $\pi^0$ events, but the FZ makes the effect much smaller. 
Also the charge exchange reaction (the third row) has a significant impact on the final states. 
It is clear that the NC $\pi^0$ production measurement is a very 
good test for the FSI models in MC event generators.

\begin{table}
\begin{tabular}{l || c | c | c}
 Channel & K2K & MB $\nu$ & MB $\bar\nu$ \\ \hline
 $1\pi^0 \rightarrow 1\pi^0$ & 93.1\% (84.5\%) & 93.0\% (88.3\%) & 94.8\% (92.4\%) \\
 $\mbox{no } \pi \rightarrow 1\pi^0$ & 2.0\% (3.2\%) & 1.8\% (2.4\%) & 1.2\% (1.6\%) \\
 $\mbox{other } \pi \rightarrow 1\pi^0$ & 3.7\% (6.8\%) & 4.2\% (5.8\%) & 3.2\% (3.9\%) \\
 $\mbox{more } \pi \rightarrow 1\pi^0$ & 1.2\% (5.5\%) & 1.0\% (3.5\%) & 0.7\% (2.1\%) \\
\end{tabular}
\caption{Origin of the events with $1\pi^0$ in the final state. Values in brackets refer to results without FT. }
\label{tab: ncpianal1}
\end{table}

\begin{table}
\begin{tabular}{l || c | c | c}
 Channel & K2K & MB $\nu$& MB $\bar\nu$ \\ \hline
 $1\pi^0 \rightarrow 1\pi^0$ & 81.6\% (64.0\%) & 79.1\% (66.9\%) & 83.0\% (74.5\%) \\
 $1\pi^0 \rightarrow \mbox{no } \pi$ & 5.9\% (19.3\%) & 7.2\% (19.2\%) & 6.4\% (15.9\%) \\
 $1\pi^0 \rightarrow \mbox{other } \pi$ & 10.1\% (11.0\%) & 10.2\% (10.1\%) & 9.6\% (7.8\%) \\
 $1\pi^0 \rightarrow \mbox{more } \pi$ & 2.4\% (5.7\%)& 2.0\% (3.7\%) & 1.0\% (1.8\%) \\
\end{tabular}
\caption{Impact of FSI efects on the events with $1\pi^0$ in the primary interaction. Values in brackets refer to the results without FT.}
\label{tab: ncpianal2}
\end{table}

Table \ref{tab: sb} shows the composition of the $\pi^0$ signal in the SciB experiment as it is understood by NuWro. 
The second column contains the values reported by the SciB collaboration obtained from the MC they used in the data analysis (NEUT). 	

\begin{table}
\begin{tabular}{c||c|c|c}
 Channel & SciB MC & NuWro (no FT) & NuWro (FT) \\ \hline
 $1\pi^0$ & 85\% & 80\% & 82\% \\ \hline
 $1\pi^0$ + charged $\pi$ & 11\% & 16\% & 14\% \\ \hline
 $2\pi^0$ & 4\% & 4\% & 4\%
\end{tabular}
\caption{Predictions of contribution to $\pi^0$ channel from SciBooNE MC and NuWro}
\label{tab: sb}
\end{table}

\begin{table}
\begin{tabular}{c|c|c}
NC$\pi^0$/CC & K2K & SB \\ \hline
Data & 0.064$\pm$0.008 & 0.077 $\pm$ 0.010 \\
NuWro (without FZ) & 0.070 & 0.071 \\
NuWro (with FZ) & 0.079 & 0.077 \\
\end{tabular}
\caption{NC$\pi^0$/CC ratio}
\label{tab: ratio}
\end{table}

K2K and SciBooNE did not publish the normalized differential cross section. 
However, flux averaged ratios of NC$\pi^0$ to 
total CC cross sections were given. In the Table \ref{tab: ratio} we compare both values 
with the NuWro results.

Figures \ref{fig: k2k} - \ref{fig: mba} show the data and NuWro predictions  
for $\pi^0$ momentum distribution in various experiments.

In the case of the normalized cross section the main effect of the introduction of the FZ is the increase of 
the cross section in the pion absorption peak region. 
The effect can be estimated to be $10-15\%$. 
In the case of the K2K measurement the use  of the FZ also moves the peak of the pion momentum distribution to 
larger values by about $50$~MeV/c resulting in much better agreement with the data.

Both MB and SciB experiments provide distributions of events versus the cosine of the angle between the neutrino and $\pi^0$ momenta.
Figs \ref{fig: mb_angle} -- \ref{fig: sb_angle} show pions angular distributions together with the NuWro predictions. We focus on the 
backward directions because we expect an important impact from FZ effects in this kinematical region. 

Figs \ref{fig: mb_angle} and \ref{fig: mba_angle} show that the FZ increases the $\pi^0$ production in the backward directions but the effect 
is rather small. The reason is that at lower neutrino energies there are many backward moving $\pi^0$'s even without FSI effects. 
We checked that only for larger $Q^2$ values the FSI become the main {\it source} of $\pi^0$'s and 
using the FZ reduces their number.

In the case of SciB experiment the NuWro results are normalized 
to the number $\pi^0$ predicted to be in the data.
In this case using the FZ makes the absolute number of backward moving $\pi^0$'s little larger, but the effect can hardly be seen.

%%%%%%%%%%%%%%%%%%%%%%%%%%%%%%
%%%%%%%%%%%%%%%%%%%%%%%%%%%%%
\section{Conclusions}
%%%%%%%%%%%%%%%%%%%%%%%%%%%%%
%%%%%%%%%%%%%%%%%%%%%%%%%%%%%

Any comparison to recent NC $\pi^0$ production data requires a computational tool capable of modeling several dynamical mechanisms 
for neutrino-nucleon interaction as well as the FSI effects. 
The NuWro MC event generator has all the required
physical models implemented and we have demonstrated that
it reproduces the experimental results quite well. An important ingredient of the NuWro FSI model is the FZ mechanism which even at
relatively small neutrino energies typical for K2K, MB and SciB experiments leads to observable effects on the $\pi^0$'s in the final state. 
We hope that our results will be useful for better evaluation of the systematic error coming from NC $\pi^0$ production in 
neutrino oscillation experiments like T2K.

%%%%%%%%%%%%%%%%%%%%%%%%%%%%%%%%%%%%%

\begin{acknowledgments}
The authors were partially supported by the grants: N N202 368439 and 2011/01/M/ST2/02578.
\end{acknowledgments}

%%%%%%%%%%%%%%%%%%%%%%%%%%%%%%
%%%%%%%%%%%%%%%%%%%%%%%%%%%%%%

\end{document}